\documentclass[fleqn,usenatbib]{mnras}  

\usepackage{amsmath}
\usepackage{graphicx}
\usepackage{natbib}
\usepackage{color}
\usepackage{longtable}

\usepackage[T1]{fontenc}
\usepackage{ae,aecompl}

\usepackage[normalem]{ulem}

\usepackage{tabularx}
\usepackage{comment}

\usepackage{amssymb}	
\usepackage{epsfig}
\usepackage{upgreek}
\usepackage{mathptmx}
\usepackage[utf8]{inputenc}
\usepackage{dcolumn}
\usepackage{enumerate}
\usepackage{longtable,lscape}
\usepackage{grffile}
\usepackage{array,float}
\usepackage{multirow}
\usepackage{lscape}
\usepackage{xargs}
\usepackage{float}

\usepackage{txfonts}
\include{definitions}

      \def\new#1 {{\bf #1 }}
      \def\cut#1 {\sout{#1} }

\def\kms {$\mathrm{km\,s^{-1}}$} 
\def\degr {\hbox{$^\circ$}}
\def\percc {$\mathrm{cm^{-3}}$} 
\def\cmsq  {$\hbox{{\rm cm}}^{-2}$}    

\def\HII {H{\sc ii}} 
\def\NTH {$\mathrm{N_2H^{+}}$} 
\def\HCOP {$\mathrm{HCO^{+}}$} 

\def\simgreat{\mathbin{\lower 3pt\hbox
     {$\rlap{\raise 5pt\hbox{$\char'076$}}\mathchar"7218$}}}
\def\simless{\mathbin{\lower 3pt\hbox
     {$\rlap{\raise 5pt\hbox{$\char'074$}}\mathchar"7218$}}}

\newcolumntype{d}[1]{D{.}{\cdot}{#1}}

\newcolumntype{.}{D{.}{.}{-1}}

\newcommand{\msun}{M$_\odot$}

\newcommand{\rgc}{$R_{\rm gc}$}

\newcommand{\mum}{$\mu$m}

\newcommand{\hii}{H{\sc ii}}

\title[Infall and Outflow towards Starless Clumps]{Infall and Outflow Towards High-mass Starless Clump Candidates}

\author[T. Pillai et al.]{T.\,G.\,S.\,Pillai,$^{1}$\thanks{E-mail: thushara@mit.edu}
J.\,S.\,Urquhart,$^{2}$\thanks{E-mail: j.s.urquhart@kent.ac.uk}
S.\,Leurini,$^{3}$
Q.\,Zhang,$^{4}$
A.\,Traficante,$^{5}$
D.\,Colombo,$^{6}$\newauthor
K.\,Wang,$^{7}$ 
L.\,Gomez,$^{8}$
F.\,Wyrowski$^{6}$
\\
$^{1}$ Institute for Astrophysical Research, Boston University, 725 Commonwealth Avenue, Boston, MA 02215, USA \\
$^{2}$ Centre for Astrophysics and Planetary Science, University of Kent, Canterbury, CT2\,7NH, UK\\
$^{3}$ INAF – Osservatorio Astronomico di Cagliari, Via della Scienza 5, I-09047 Selargius (CA), Italy\\
$^{4}$  Center for Astrophysics, Harvard \& Smithsonian, 60 Garden Street, Cambridge, MA 02138, USA \\ 
$^{5}$ IAPS - INAF, via Fosso del Cavaliere, 100, I-00133 Roma, Italy \\ 
$^{6}$ Max-Planck-Institut f\"ur Radioastronomie, Auf dem Hügel 69, 53121 Bonn, Germany \\ 
$^{7}$ Kavli Institute for Astronomy and Astrophysics, Peking University, 5 Yiheyuan Road, Haidian District, Beijing 100871, China \\
$^{8}$ Joint Alma Observatory, Alonso de C\'ordova 3107, Vitacura, Santiago, Chile
}

\date{Accepted XXX. Received YYY; in original form ZZZ}

\pubyear{2022}

\begin{document}
\label{firstpage}
\pagerange{\pageref{firstpage}--\pageref{lastpage}}
\maketitle

\begin{abstract}
The evolutionary sequence for high-mass star formation starts with massive starless clumps that go on to form protostellar, young stellar objects and then compact \hii\ regions. While there are many examples of the three later stages, the very early stages have proved to be elusive.  We follow-up a sample of 110 mid-infrared dark clumps selected from the ATLASGAL catalogue with the IRAM telescope in an effort to identify a robust sample of massive starless clumps. We have used the \HCOP\ and HNC (1-0) transitions to identify clumps associated with infall motion and the SiO (2-1) transition to identity outflow candidates.  We have found blue asymmetric line profile in 65\,per\,cent of the sample, and have measured the infall velocities and mass infall rates (0.6 -- $36 \times 10^{-3}$\,M$_\odot$\,yr$^{-1}$) for 33 of these clumps. We find a trend for the mass infall rate decreasing with an increase of bolometric luminosity to clump mass i.e. star formation within the clumps evolves. Using the SiO 2-1 line, we have identified  good outflow candidates. Combining the infall and outflow tracers reveals that 67\,per\,cent of quiescent clumps are already undergoing gravitational collapse or are associated with star formation; these clumps provide us with our best opportunity to determined the initial conditions and study the earliest stages of massive star formation. Finally, we provide an overview of a systematic high-resolution ALMA study of quiescent clumps selected  that allows us to develop a detailed understanding of earliest stages and their subsequent evolution.
\end{abstract}




\section{Introduction}

Massive stars are born within giant molecular cloud (GMC) complexes and reach the main sequence on very short timescales such that finding true massive protostars remains a challenge. In particular, finding and characterizing the massive clumps prior to the onset of star formation is still one of the outstanding problems in star formation research \citep{tan2014}. 

The vast majority of molecular gas resides within the Solar Circle (i.e. Galactocentric distances (\rgc) $< 8.35$\,kpc) and given the tight correlation found between molecular gas and star formation within the Galaxy (e.g. \citealt{heiderman2010, lada2010}) and in more distance galaxies (e.g. \citealt{wu2004}) this is where star formation is also expected to be concentrated. The inner part of the Galaxy that hosts most of the massive stars and clusters corresponds to a Galactic longitude ($\ell$) range of $\pm 60$\,\degr, and this is an obvious region to focus efforts to develop a deeper understanding of massive star formation. 

Depending on the source of mass accretion on to the most massive star, models of high-mass star formation  fall into two broad categories: clump-fed accretion and core accretion.  The former models are characterized by gas assembly through either global clump infall  or coherent gas flows that results in the formation of high-mass stars in clusters. In certain clump-fed models, clump fragmentation produces Jeans-mass cores that proceed to accrete mass through gas infall from their larger environment. The low-mass cores that are closer to the centre of the evolving gravitational potential preferentially gain more mass to end up forming the highest mass stars in the cluster. Competitive accretion models \citep{bonnell2004, bonnell2006, smith2009, wang2010}, global hierarchical collapse models \citep{semadeni19} and a more recent inertial-inflow model \citep{padoan2020} fall into the  "clump-fed" category.

The turbulent core accretion model \citep{mckee02} posits that a high-mass star or cluster forms out of an unfragmented dense and high-mass starless core. It therefore treats the formation of high-mass stars in isolation rather than as part of cluster formation. Global infall in this scenario is slow \citep{tan2006}, and is not a major reservoir for the high-mass starless core. High-mass starless cores are structures in near-virial equilibrium with a one-on-one mapping to high-mass stars and are a prerequisite for this model \citep{rosen2019}.

Fundamental predictions of the above models needs to be tested on an unbiased sample of high-mass star forming clumps. The APEX Telescope Large Area Survey of the Galaxy (ATLASGAL; \citealt{schuller2009_full}) has produced the first such unbiased  870\,\mum\ dust survey of the inner Galactic plane. This survey is sensitive to clump masses of $\sim 1000$\,\msun\ across the inner Galactic disk (\citealt{urquhart2014_atlas}) and is therefore likely to include all massive star forming clumps located within the inner Galactic disc. Furthermore, since the thermal dust continuum emission at 870\,\mum\ is optically thin, it is sensitive to both cold and warm dust and is not biased to a particular evolutionary stage. For these reasons this survey is the ideal starting point to search for the massive starless cold clumps that are the progenitors to high-mass protostars and YSOs and ultracompact (UC) \hii\ regions, which have proved to be elusive.

\subsection{Overview of the ATLASGAL Catalogue}
\label{sect:atlasgal_overview}

ATLASGAL covers 420 square degrees of the inner Galactic disc  ($-60$\degr$\le \ell \le +60$\degr,
$-1.5$\degr$\le b \le +1.5$\degr; \citealt{schuller2009_full}) and was conducted with the 12-m Atacama Pathfinder Experiment (APEX) telescope (\citealt{gusten2006}). The resulting dust maps have been used to identify $\sim$10\,000 dense molecular clumps primarily located within the Solar circle (\citealt{contreras2013, urquhart2014_csc,csengeri2014}) and includes samples of sources in all of the early evolutionary stages associated with high-mass star formation (\citealt{konig2017}). Dedicated follow-up observations (\citealt{wienen2012,giannetti2014,wienen2015, csengeri2016_sio, kim2017, kim2018, wienen2018,tang2018, navarete2019,urquhart2019, kim2020}) and complementary surveys (e.g. WISE (\citealt{cutri2012}), HiGAL (\citealt{Molinari2010}), CORNISH (\citealt{hoare2012, purcell2013, irabor2023}), MALT90 (\citealt{jackson2013,guzman2015, rathborne2016, contreras2017}) have been used to characterise the physical properties of these clumps (\citealt{urquhart2014_atlas, urquhart2018}), map their Galactic distribution and investigate the evolutionary sequence for high-mass star formation (\citealt{urquhart2022}). The ATLASGAL catalogue is complete to all potential massive star forming clumps in the inner Galaxy and provides robust physical properties and is therefore an ideal starting point for more focused high-resolution studies.

Our current understanding of massive stars distinguishes the actively accreting high luminosity protostars in clusters with no cm continuum (high-mass protostars or clusters) from those with cm continuum (hypercompact and ultracompact \HII\ regions). In the framework of the legacy project aimed at a complete characterization of massive stars across the evolutionary ladder,  ATLASGAL has delivered a comprehensive sample of molecular clumps containing (i) massive and young protostars within 4.5 kpc \citep{csengeri2017_spark}, (ii) a more evolved phase characterized by 6.7 GHz methanol maser emission \citep{urquhart2013_methanol, billington2019_meth}, and finally (iii) compact and ultracompact \HII\ regions \citep{urquhart2013_cornish}. The initial evolutionary stage before the onset of protostellar collapse, where the young clumps are dense, cold and starless, is a crucial phase yet to be characterized for the ATLASGAL project.

In this paper we report on the results of a molecular line survey conducted with the IRAM telescope towards a sample of 70\,\mum\ dark clumps that are assumed to be high-mass starless clumps. We focus specifically on the transitions that trace infall and outflow to distinguish between genuinely starless clumps and those showing signs of very early star formation. The structure of the paper is as follows: In Sect.\,\ref{sect:obs} we describe the source selection criteria, provide an overview of the IRAM 30-m spectral line observations and the data reduction processes. In Sect.\,\ref{sect:results} we present the detection statistics and investigate their bulk properties (e.g. outflows, infall motions). In Sect.\,\ref{sect:discussion} we discuss the nature of the sample and their properties and describe our motivation for  high angular resolution follow up efforts with ALMA. In Sect.\,\ref{sect:conclusions} we provide a summary of the work presented here and provide an outline of the next papers in this series.

\section{Observations}
\label{sect:obs}

\subsection{Source selection}
\label{sect:source_selection}

We have initiated a search for the coldest dust cores in the inner Galactic plane based on the ATLASGAL data (\citealt{schuller2009_full}). Cold and dense starless cores are distinguished from protostellar cores based on the lack of IR emission, particularly at 24\,\mum\ where protostellar activity would otherwise warm up the dust. 

We began by cross matching the whole ATLASGAL catalogue (\citealt{contreras2013, urquhart2014_csc}) with the WISE catalogue and excluded all clumps with a mid-infrared counterpart within 20 arcsec. In order to select only massive clumps, we estimate the clump masses, using the catalogue distances (\citealt{urquhart2018}) and assuming a dust temperature of 20\,K, only selecting those that exceed the empirical mass-size threshold for high-mass star formation (\citealt{kauffmann2010c}; $m(r)/m_{\rm lim}(r)>2$, where $m_{\rm lim}(r)  = 870 M_{\odot}[r/\text{pc}]^{1.33}$). While our adopted dust temperature of 20\,K would be appropriate for the majority of the sample which is in a cold phase, slightly higher temperatures may be expected for the more evolved sources in the sample \citep{urquhart2022} and could introduce a small bias well within a factor of two in the mass estimate. Finally, we also exclude any sources more than 1\degr\ from the Galactic mid-plane (i.e. $|b| > 1$\degr). These criteria produced a sample of 188 clumps, 110 of which are observable from the IRAM\,30-m telescope, which is located at Pico Veleta in the Spanish Sierra Nevada. 

\begin{figure}
    \centering
  
    \includegraphics[width=.45\textwidth, trim= 20 50 0 50,clip]{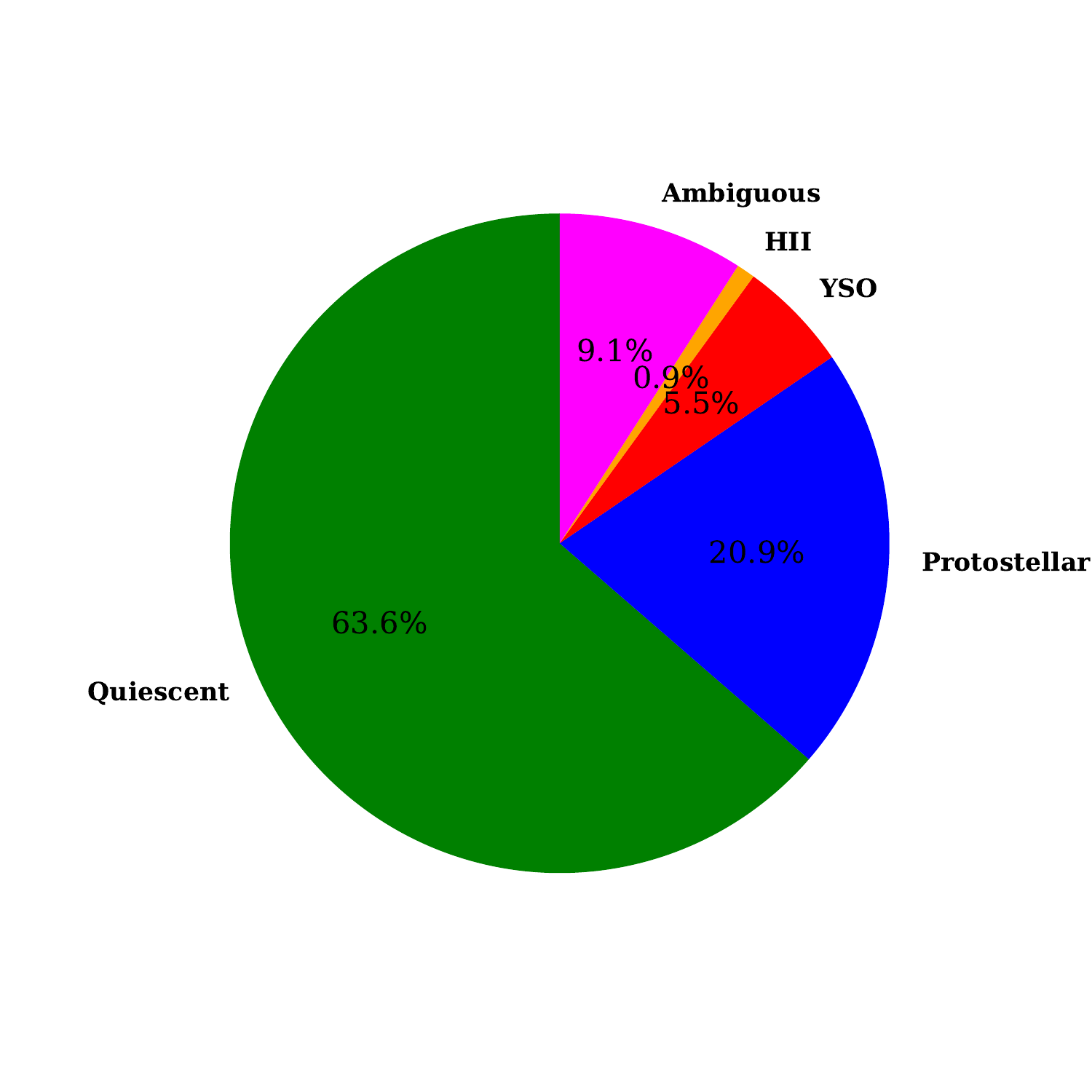}
    
    
\caption{Pie chart showing the fraction of clumps in one of four evolutionary stages based on \citep{urquhart2022} or classified as ambiguous in cases where a definitive classification was not possible. The total sample consists of 110 clumps.}

\label{fig:pie_chart}
\end{figure}

The original source selection was done early in the characterisation of the ATLASGAL catalogue when distances and HiGAL 70\,\mum\ images were not readily available. Since that time, despite our best efforts, a modest portion of the sample has since been found to be associated with star formation activity. In Figure\,\ref{fig:pie_chart} we show the distribution of this sample as a function of the four evolutionary stages described by  \citet{urquhart2022}; these are, in sequences,  quiescent, protostellar, young stellar object (YSO) and \hii\ region stages. Clumps where a definitive classification into one of these four stages cannot be made are classified as ambiguous. Although the sample contains some more evolved stages it is clear from Fig.\,\ref{fig:pie_chart} that the vast majority are in a very early stage in their evolution ($\sim$85\,per\,cent are classified as protostellar or pre-protostellar stage) and almost two-thirds show no signs that star formation has begun. 


\subsection{IRAM 30-m observations and data reduction}
\label{sect:obs_setup}

A sub-sample of 110 targets were observed with the IRAM 30\,m telescope in March 2016 (IRAM project code: 134-15) under varying weather conditions (1.3-3\,mm precipitable water vapor).  The EMIR receiver was used allowing these observations to cover the 86-93\,GHz frequency range in one setup in dual polarization \citep{carter2012}.  The effective spectral resolution is $\sim 0.6$\,km\,s$^{-1}$.  The observations were conducted in total power position-switching mode with 10 mins per position. We used Herschel 250\,$\mu$m data to identify clean off-positions located close to our target sources. However, there was still contamination in lower density tracers such as \HCOP\ in the off-positions towards a few sources (e.g. AGAL031.464+00.186, AGAL022.634+00.112, AGAL022.304-00.629) and in some cases these sources needed to be excluded from some parts of the analysis if it significantly affects the line profile and biases the line fitting. Pointing and focus measurements were checked regularly during the observing runs.

These observations have a frequency coverage of 8\,GHz and many spectral lines ($\sim$20), however, in this paper we focus on specific lines that are associated with direct or indirect signposts of clump collapse and star formation activity,  namely, the 1-0 transitions of HNC and \HCOP\ and their rarer optically thin isotopomers, and 2-1 transition of SiO; these are often used to identify infall and outflow motions in dense clumps (e.g. \citealt{traficante2017}). In addition \NTH\ 1-0 is used as a dense gas tracer. The CLASS package  within the GILDAS software\footnote{https://www.iram.fr/IRAMFR/GILDAS} was used for data processing. This involved extracting the specific spectral lines studied here, defining a masking window around the lines and subtracting a baseline of order 3. The spectra are reported here in antenna temperature units ($T_{\rm A}^{\star}$). The mean system temperature over the observing session was $\sim 90$\,K within a range of 80-110\,K. The corresponding mean rms is estimated to be $\sim 20$\,mK per $\sim 0.6$\,\kms\ spectral channel.

\section{Results}
\label{sect:results}

\begin{figure*}
  \centering    
  
  \includegraphics[width=.49\textwidth]{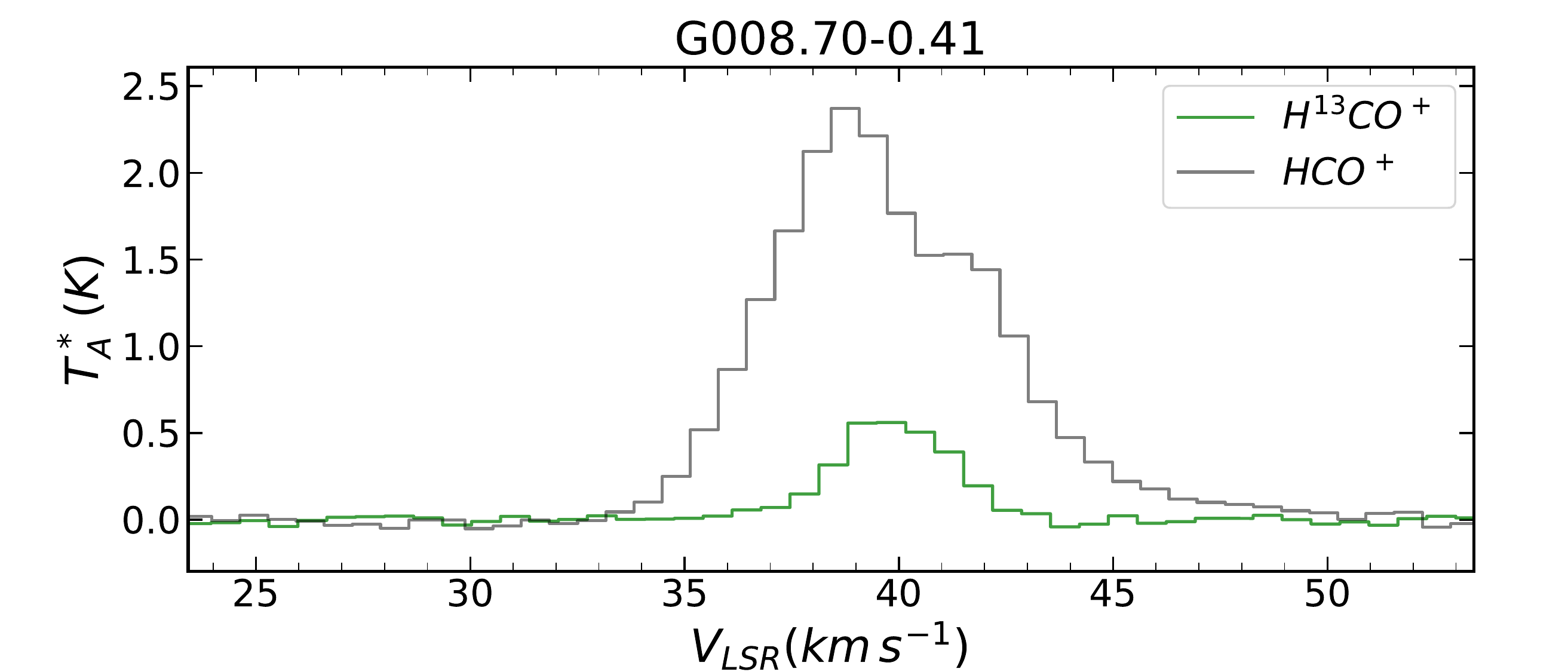}
  \includegraphics[width=.49\textwidth]{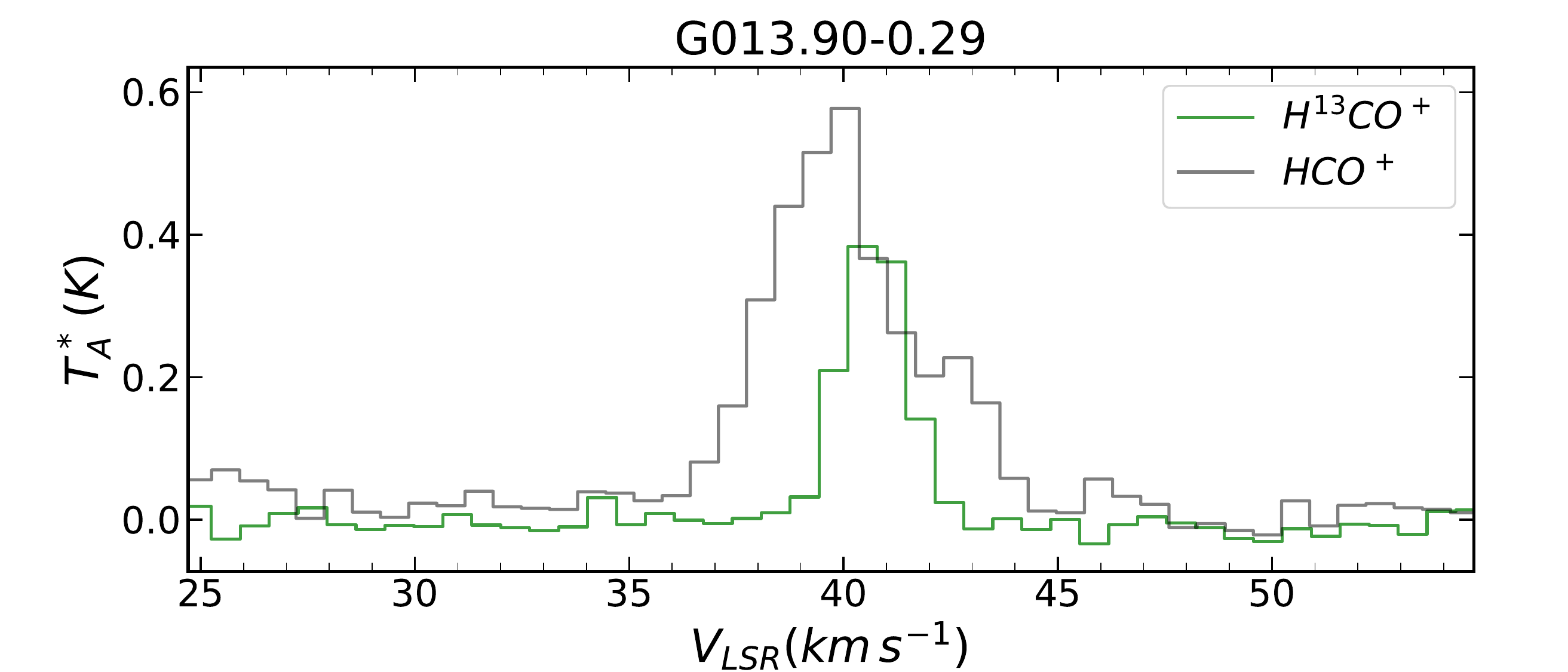}\\
  \includegraphics[width=.49\textwidth]{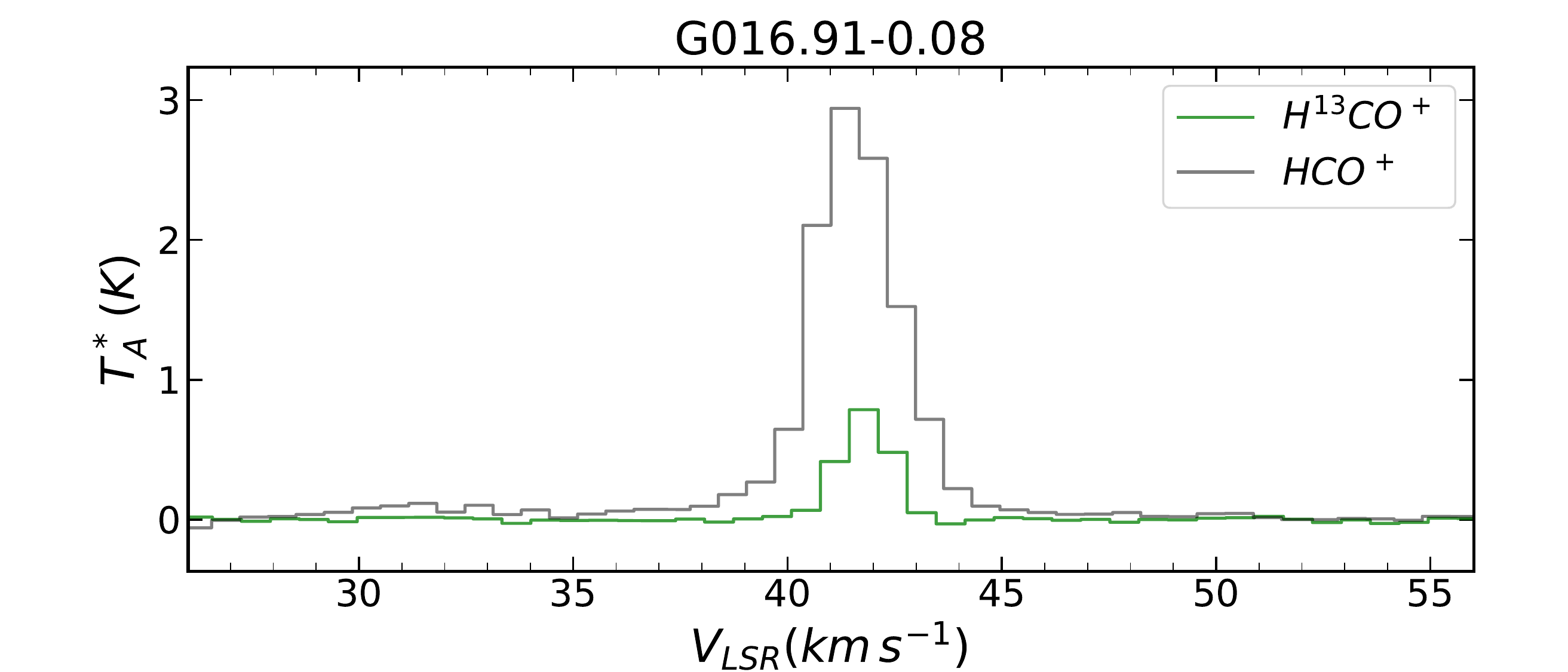}
  \includegraphics[width=.49\textwidth]{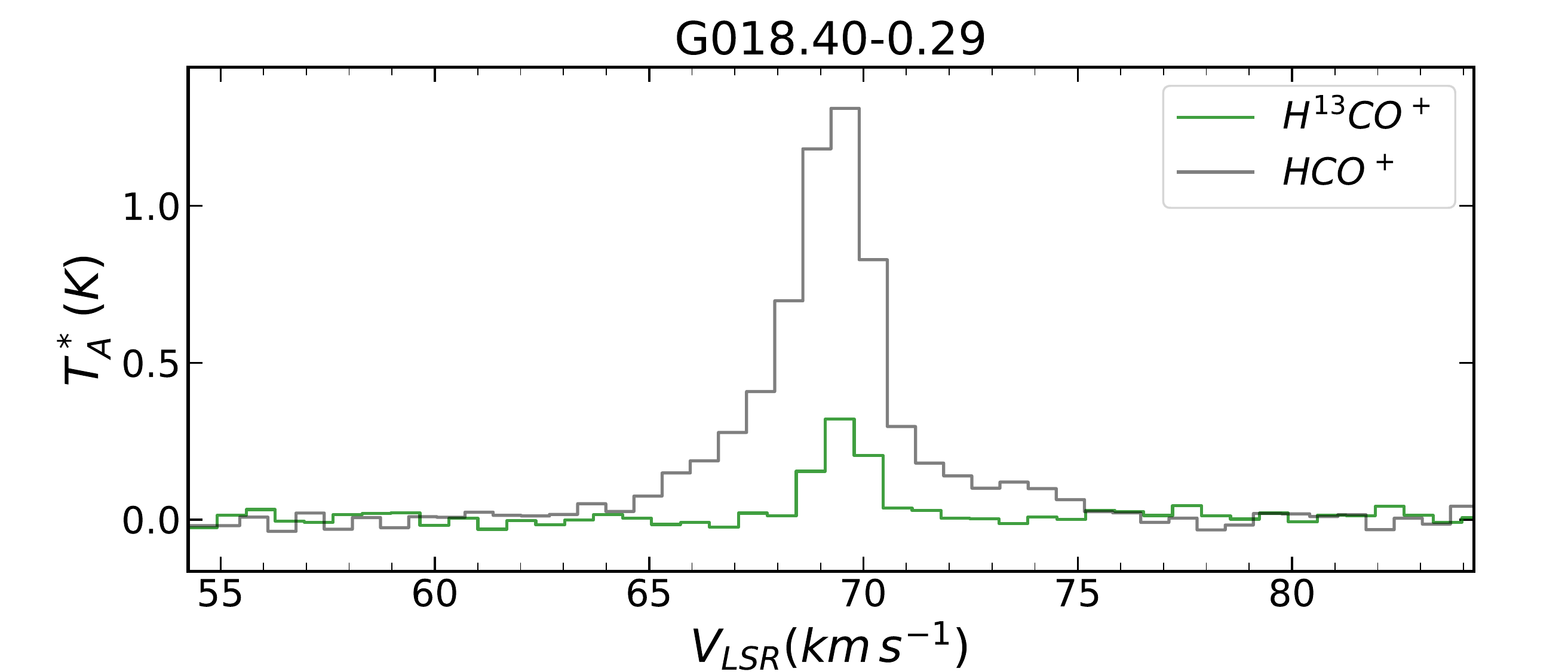}\\
\includegraphics[width=.49\textwidth]{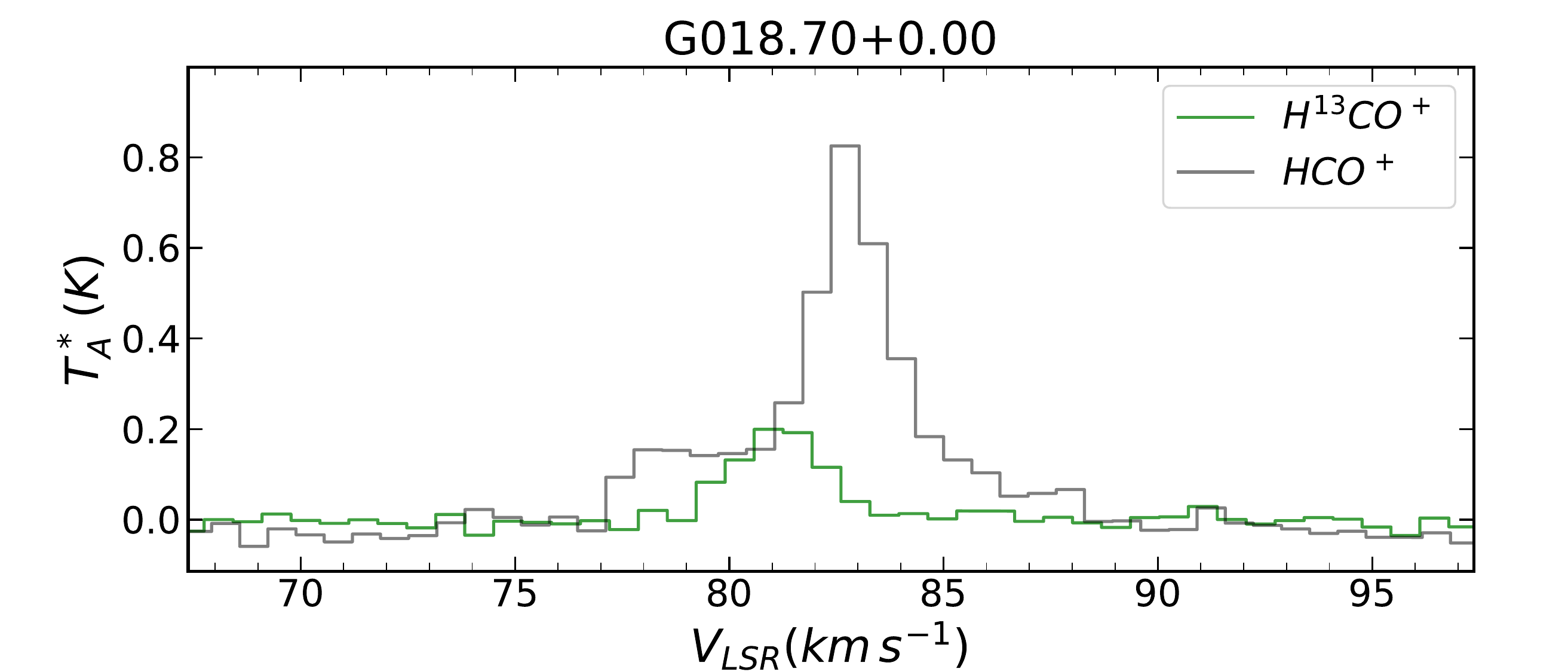}
\includegraphics[width=.49\textwidth]{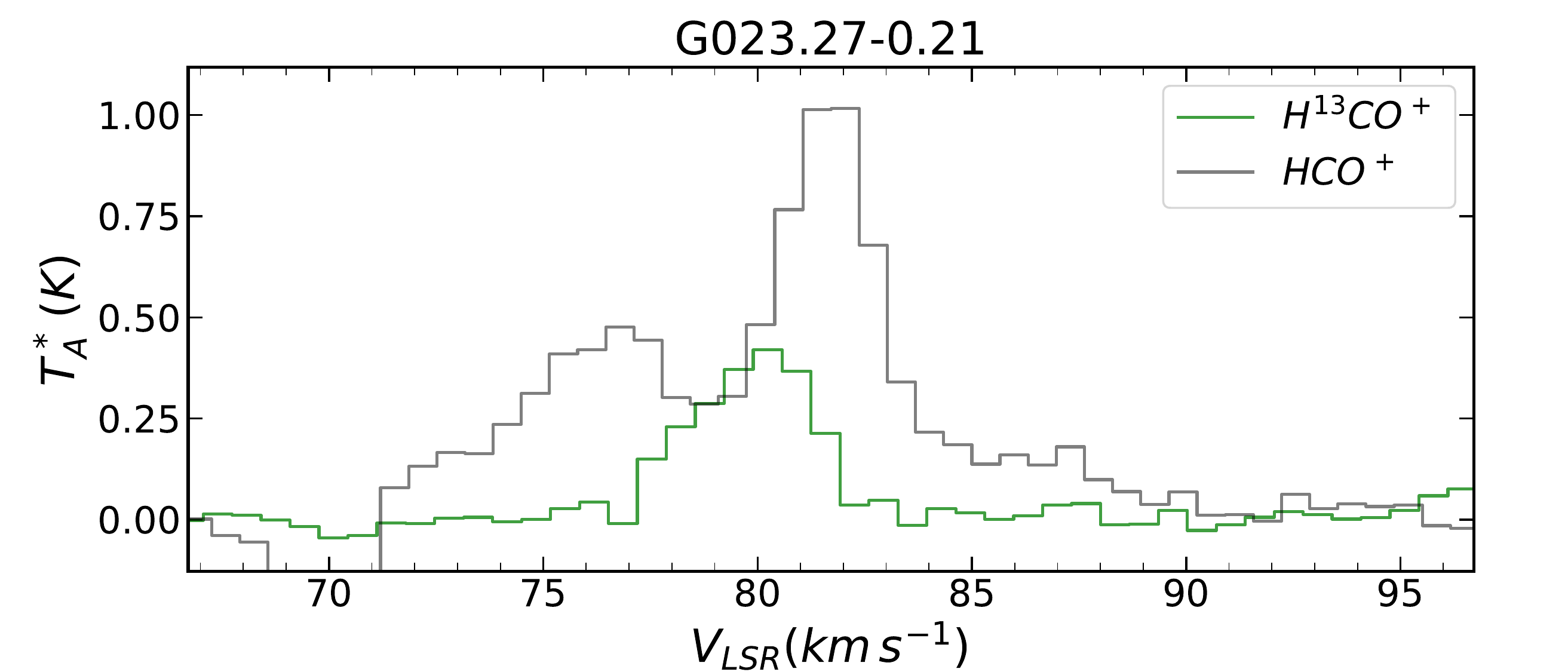}\\

\caption{Examples of  \HCOP\,(1-0) and H$^{13}$CO$^+$ (1-0) spectra showing blue asymmetric profiles indicative of infall motion. The upper panels show examples of blue-shifted asymmetries, the middle panels show examples where no asymmetries are seen, the lower panels show examples of red-shifted asymmetries.}  \label{fig:hcop_spec}
\end{figure*}

\begin{figure}

 \includegraphics[width=.49\textwidth]{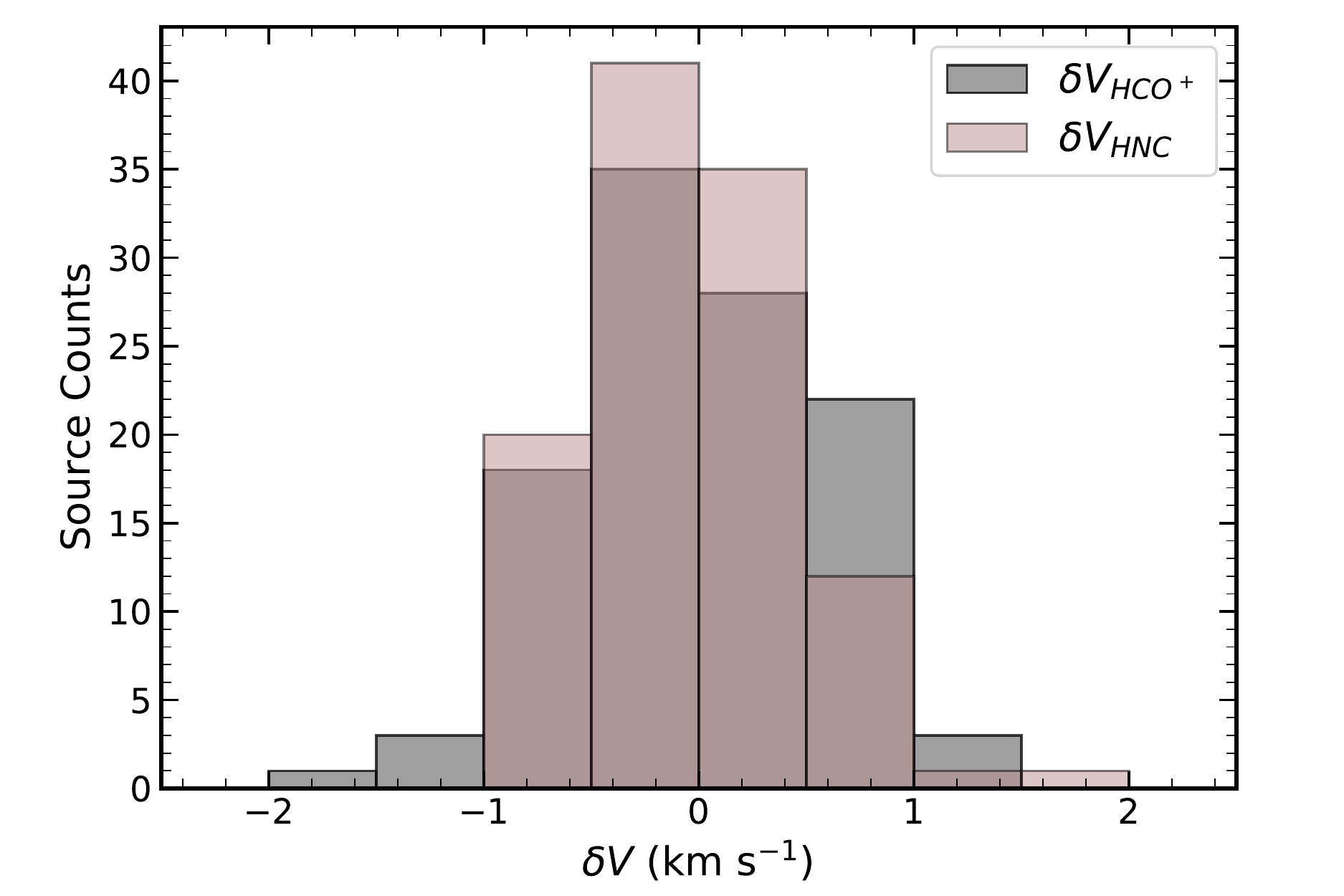}
 \includegraphics[width=.49\textwidth]{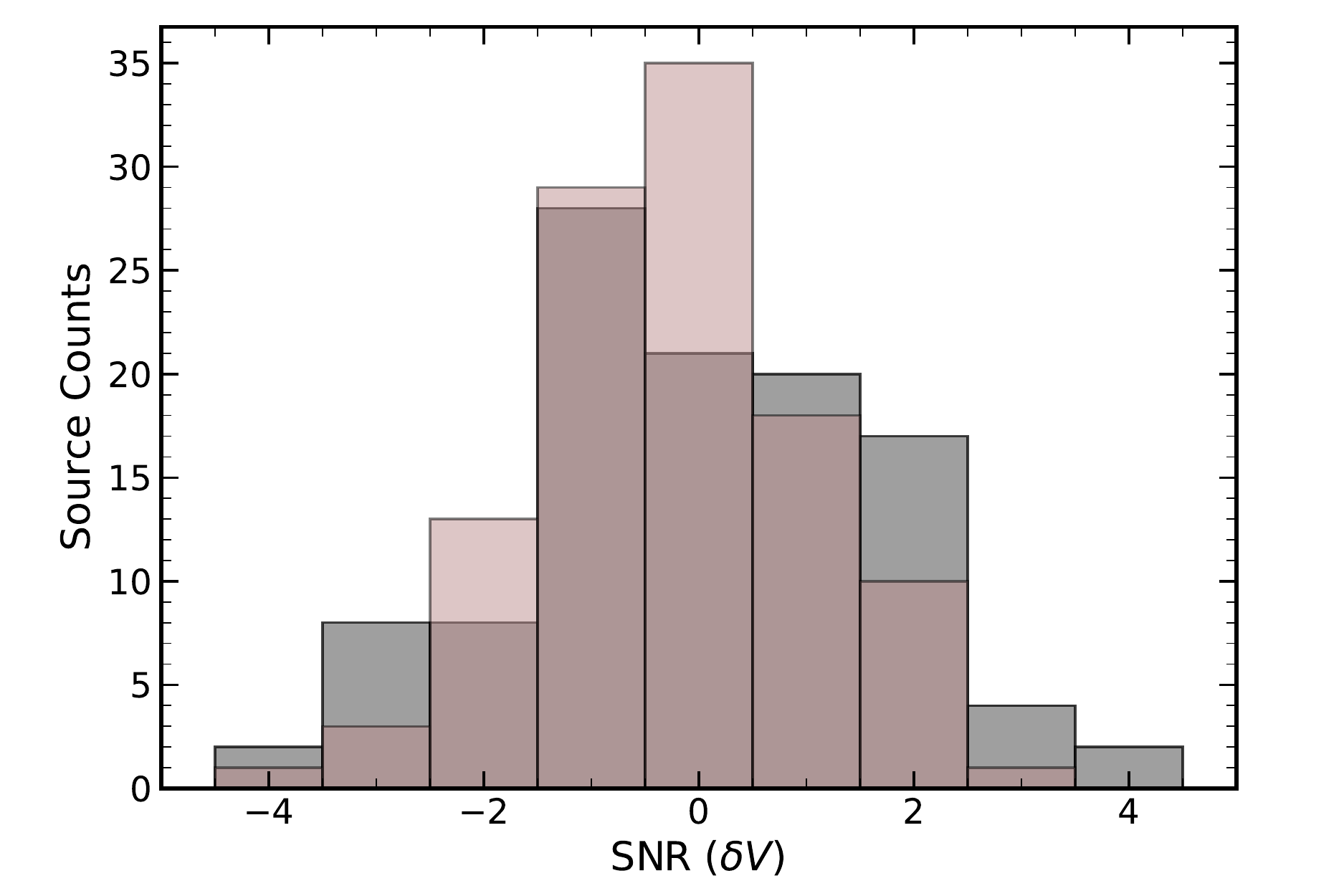}
\caption{Top Panel: Distribution of the dimensionless parameter $\delta V$ for the \HCOP\ and HNC transitions. Positive and negative values indicate a red and blue asymmetries respectively. Note more sources have negative velocity differences than positive values. The bin size is 0.5\,km\,s$^{-1}$. Bottom Panel: Distribution of the SNR on $\delta V$ for the \HCOP\ and HNC transitions. }    \label{fig:asymmetries_hist}
\end{figure}

\subsection{HCO$^+$ and HNC line asymmetries}

 Infall motions in a dense clump are typically diagnosed through asymmetric profiles in optically thick lines. For a spherically symmetric cloud with increasing densities and/or temperatures toward its center, the corresponding excitation temperature increases toward the inner parts of the cloud as well. An optically-thin spectral line should manifest as a symmetric profile since photons from the emitting gas along the entire line of sight contribute to the observed line intensities without being absorbed. When the line emission becomes optically thick, the layer corresponding to an optical depth of unity occurs at different parts in the clouds: the red-shifted emission in the observed line originates from material in front of the clump, and becomes optically thick at the clump outer layer. On the other hand, the blue-shifted emission of the line arises from the rear end, and the optically thick part is closer to the center. Since the excitation temperature increases inward in a collapsing clump, this difference causes the blue-shifted portion of the line to be stronger than the red-shifted counterparts.  This blue-and-red asymmetry in an optically thick line in combination with an optically thin line has been  used widely as an infall diagnostic in dense molecular clumps \citep{mardones1997, evans2003, fuller2005, evans2015, jackson2019}.  Figure\,\ref{fig:hcop_spec} show examples from our IRAM survey that show that the peak of the optically thin line (H$^{13}$CO$^+$ 1-0) lies between the double peaks of the thick line (\HCOP\,1-0). Assuming a typical $^{12}$C/$^{13}$C ratio of 50 \citep{wilson1994} and at least a 3\,$\sigma$ detection of H$^{13}$CO$^+$ 1-0, one would naively expect a S/N of $>150$ for the \HCOP\ 1-0 line, not observed in our data. The significant detection ($\gg 3\,\sigma$) of H$^{13}$CO$^+$  1-0 line towards all targets thus suggests that the \HCOP\ 1-0 is optically thick.
 
 Note however that the sense of the line asymmetry is predicated on the assumption that the excitation temperature increases inward. Should it decrease inward, the emergent optically thick line will reverse its sense of asymmetry,  exhibiting a stronger red-shifted peak than the blue-shifted peak. Additionally, certain molecular species (e.g. \HCOP) might be enhanced in  molecular outflows \citep{arce2006,cyganowski2011} as well and potentially obscure an infall signature.

We first identify sources with significant line asymmetries using the dimensionless parameter $\delta V$ defined by \citealt{mardones1997}:

\[
\delta V = \frac{V_{\rm thick} - V_{\rm thin}}{\Delta V_{\rm thin}}
\]

\noindent where $V_{\rm thick}$ and $V_{\rm thin}$ are the peak velocities of the optically thick and thin transitions, and $\Delta V_{\rm thin}$ is the FWHM line width of the optically thin transition. Dividing through by the FWHM normalises the parameter and avoids the possibility of bias from lines with different line widths. The velocities are determined simply by taking the velocity of the channel with the peak intensity while the line width of the optically thin line is determined from a Gaussian fit. This parameter is not reliable in cases where the optically thick line has two peaks with similar intensity. There are three such cases in our \HCOP\ sample (AGAL010.991$-$00.082, AGAL023.990+00.149 and AGAL024.314+00.086) and one case in our HNC sample (AGAL025.163$-$00.304); these have been excluded from this analysis.

The distributions of the asymmetries for the \HCOP\ and HNC transitions are shown in Fig.\,\ref{fig:asymmetries_hist} (top panel). These plots reveal a roughly even number of blue and red asymmetries in both transitions, with slightly higher number of blue symmetries (negative values; cf. \citealt{mardones1997}). In total, 71 clumps show a blue asymmetry in at least one of the optically thick lines, corresponding to approximately 65\,per\,cent of the sample. We note that our coarse velocity resolution could potentially obscure double peaked profiles with smaller infall motions that might prevail in the remaining sample.

We conduct a more thorough assessment of the presence of asymmetry in our sample that takes the uncertainties in line profile measurements for individual sources into account. For a given source we  first calculate $\delta{}V$ from the observed data. We then take the observed \HCOP\ and HNC spectra, add random noise as characteristic for our observed data to these. This noise spectrum is then added to our observed data. We then calculate the asymmetry parameter for each source. This insertion of random noise is repeated $10^4$ times per observed spectrum, and the asymmetry parameter is calculated in every iteration of this process. We then obtain the uncertainly of the asymmetry parameter as the standard deviation of the synthetic measurements produced by insertion of noise, $\sigma(\delta{}V)$. The resultant SNR of the asymmetry parameter, $\delta{}V/\sigma(\delta{}V)$, calculated using these newly-derived uncertainties is what is shown in Fig.\,\ref{fig:asymmetries_hist} (bottom panel) and used to create Table\,\ref{tab:line-asymmetries}. The latter reports the observed number of sources for various ranges of $\delta{}V/\sigma(\delta{}V)$, $N_{\rm{}obs}$, and the number of cases expected for a Gaussian distribution, $N_{\rm{}Gauss}$. Table\,\ref{tab:line-asymmetries} demonstrates that $N_{\rm{}obs}>N_{\rm{}Gauss}$ for a wide range of $\delta{}V/\sigma(\delta{}V)$, something also evident in Fig.\,\ref{fig:asymmetries_hist} (bottom panel). In other words, substantial asymmetries are observed much more often than consistent with pure chance as induced by noise. Line asymmetries are substantial in our sample, indicating the presence of significant relative gas motions in our targets.

We have also evaluated the line asymmetry following a new method introduced by \citet{jackson2019}. In this case the optically thin line is used to determine the systemic velocity using H$^{13}$CO$^+$ and HN$^{13}$C, and the integrated line intensities on the blue and red side of this systemic velocity are then calculated for the optically thick line, $I_{\rm{}blue}$ and $I_{\rm{}red}$. The asymmetry is then calculated as $A=(I_{\rm{}blue}-I_{\rm{}red})/(I_{\rm{}blue}+I_{\rm{}red})$. We again create $10^4$ spectra with synthetic injected noise to determine the uncertainty of $A$ for every source, as described for $\delta{}V$. Table\,\ref{tab:line-asymmetries_A} presents the results. As in the case of $\delta{}V$, we find that $N_{\rm{}obs}>N_{\rm{}Gauss}$, meaning that line asymmetries are significant.

\begin{table}
\centering
\caption{Statistical significance of non-zero line asymmetries in $\delta{}V$.\label{tab:line-asymmetries}}
\begin{tabular}{cccc}
\hline $\delta{}V/\sigma(\delta{}V)$ & $N_{\rm{}obs}({\rm{}HCO^+})$ & $N_{\rm{}obs}({\rm{}HNC})$ & $N_{\rm{}Gauss}$ \\
\hline\hline
$<-3$ & 5 & 1 & 0.15\\
$<-2$ & 14 & 9 & 2.5\\
$<-1$ & 35 & 32 & 17\\
$>+1$ & 31 & 20 & 17\\
$>+2$ & 11 & 6 & 2.5\\
$>+3$ & 4 & 1 & 0.15\\
\hline
\end{tabular}
\end{table}

\begin{table}
\centering
\caption{Statistical significance of non-zero line asymmetries in $A$.\label{tab:line-asymmetries_A}}
\begin{tabular}{cccc}
\hline $\delta{}V/\sigma(\delta{}V)$ & $N_{\rm{}obs}({\rm{}HCO^+})$ & $N_{\rm{}obs}({\rm{}HNC})$ & $N_{\rm{}Gauss}$ \\
\hline\hline
$<-3$ & 11 & 5 & 0.15\\
$<-2$ & 19 & 12 & 2.5\\
$<-1$ & 33 & 26 & 17\\
$>+1$ & 36 & 31 & 17\\
$>+2$ & 20 & 16 & 2.5\\
$>+3$ & 11 & 5 & 0.15\\
\hline
\end{tabular}
\end{table}

\begin{figure}
  \centering    
  
  \includegraphics[width=.49\textwidth]{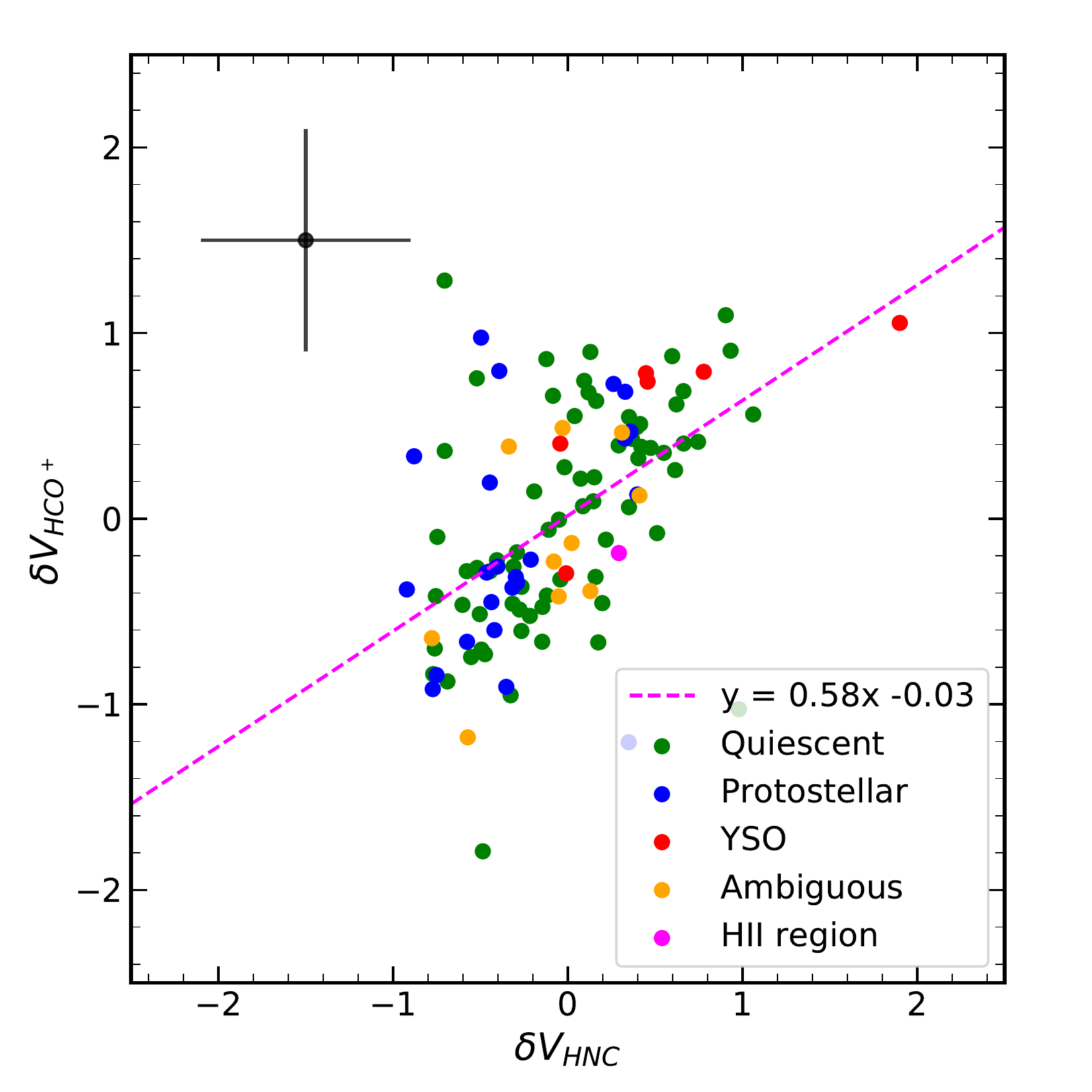} 
\caption{Distribution of asymmetries for both \HCOP\ and HNC transitions as a function of evolutionary stage. Representative uncertainties are indicated in the upper right corner; these are dominated by the channel resolution (i.e. 0.6\,\kms).  }   \label{fig:asymmetries_dist}
\end{figure}

In Figure\,\ref{fig:asymmetries_dist} we show the distribution of asymmetries as a function of the ATLASGAL source types described in Sect.\,\ref{sect:source_selection}. This plot reveals a significant correlation between the values obtained from the two transitions. Using the observed line asymmetries, we obtain a Spearman correlation coefficient $r_s = 0.6$ and a $p$-value < 0.001. We repeat the aforementioned experiment with injected artificial noise also for this correlation plot (i.e. we move individual measurements $\delta{}V_i$ by their uncertainty to obtain $10^4$ correlation diagrams for our analysis), and we find that $r_s>0.11$ for 99.73\% of all cases, equivalent to a $3\sigma$-significance. In other words, the line asymmetries seen in HNC and $\rm{}HCO^+$ are at some level correlated, and they probe similar motions. This result,  again underlines the fact that the line asymmetries are significant, as a correlation different from zero would otherwise not be found. Although there is no obvious pattern with respect to the evolutionary types we do note that nearly all of the more evolved sources tend to display red asymmetries, while the majority of the protostellar population are associated with a blue asymmetry in at least one of the two lines.

\citet{jackson2019} derive the asymmetry parameter $A$ for approximately 1000 ATLASGAL sources in their infall study and report a similar fraction of clumps associated with blue asymmetric profiles ($60.7\pm1.5$\,per\,cent) as we have found. They concluded that  the majority of their sample of high-mass star forming clumps are likely to be undergoing gravitational collapse. \citet{jackson2019} report that the asymmetry was larger for the earlier evolutionary states than for later stages. This suggests that the infall motion may be more significant in the beginning of the star formation process and decreases as the embedded pro-cluster evolves (we discuss this in more detail in Sect.\,\ref{sect:evolution_of_mass_infall_rate}). Alternatively, the infall signature in the more evolved stages is compromised by other effects, such as stronger outflows and more complicated dynamics due to feedback from already formed stars.

\subsection{Infall Analysis: HCO$^+$ and HNC}

Blue asymmetric line profiles in optically thick tracers are a powerful indicator of gravitational collapse in star-forming clumps \citep{evans1999}. However, it is not possible to measure the infall velocity for all of these due to low-signal-to-noise, contamination from the off-position and poor velocity resolution.  We have, therefore,  inspected the optically thick and thin transitions of HCO$^+$ and HNC to identify spectra where the infall signature can be reliably fitted. We then used the Python spectral line fitting module \texttt{pyspeckit} and the \texttt{Hill5} function to fit the blue and red-shifted peaks and estimate the infall velocity using the model of  \citet{de_Vries2005}. 

\begin{figure}
  \centering    
  
  \includegraphics[width=.49\textwidth]{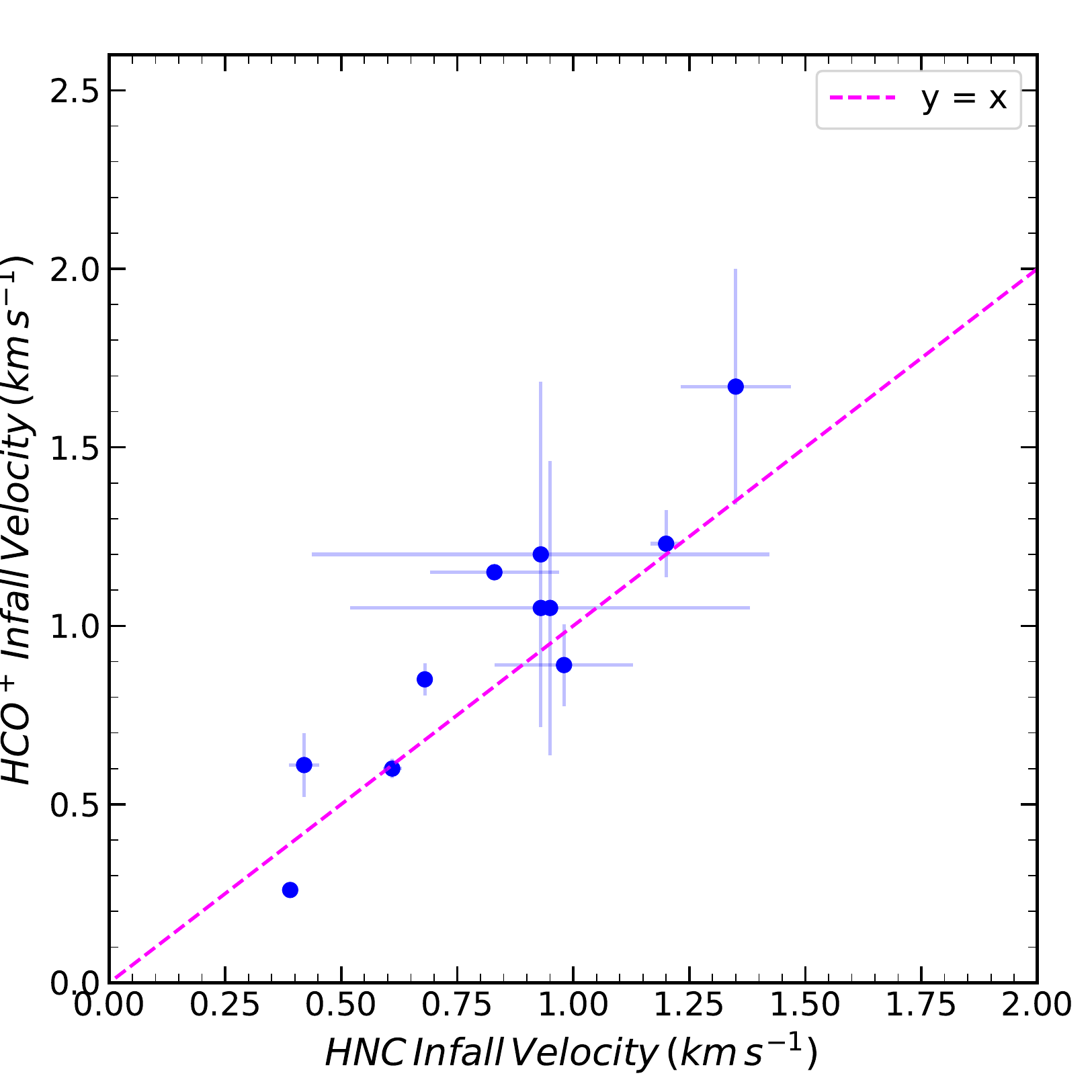} 
\caption{Comparison of the infall speeds as measured from the two optically thick lines.}   \label{fig:infall_scatter}
\end{figure}

Using this method we have been able to determine the infall velocity for 33 clumps, with 7 measurements from the HCO$^+$ transition, 15 measurements from the HNC,  and 11 measurements in both transitions. 
Eleven of these are classified as protostellar and the remaining 22 being classified as being quiescent, this corresponds to almost a third of the quiescent clumps in the sample ($\sim$32\,per\,cent). With $26\pm5.14$ detections in HNC compared to $18\pm 4.7$ detections in HCO$^+$, we also note that the 1-0 transitions of HNC appears to be more useful in detecting and quantifying the infall motion in dense clumps than  HCO$^+$. However, combining both transitions is significantly more effective than using a single transition. Enhanced abundance of HNC relative to its isomer HCN observed at low temperatures and reproduced by temperature dependence in astrochemical models might provide some explanation for the moderately higher HNC detections in our pre-dominantly young (cold) sample \citep{hacar2020}. In Figure\,\ref{fig:infall_scatter} we compare the infall speeds for the two different optically thick lines for which a measurement has been possible. In general, the agreement is very good for these eleven sources 11 with all of the values agreeing within a factor of two.

\begin{figure}
  \centering    
  
  \includegraphics[width=.49\textwidth]{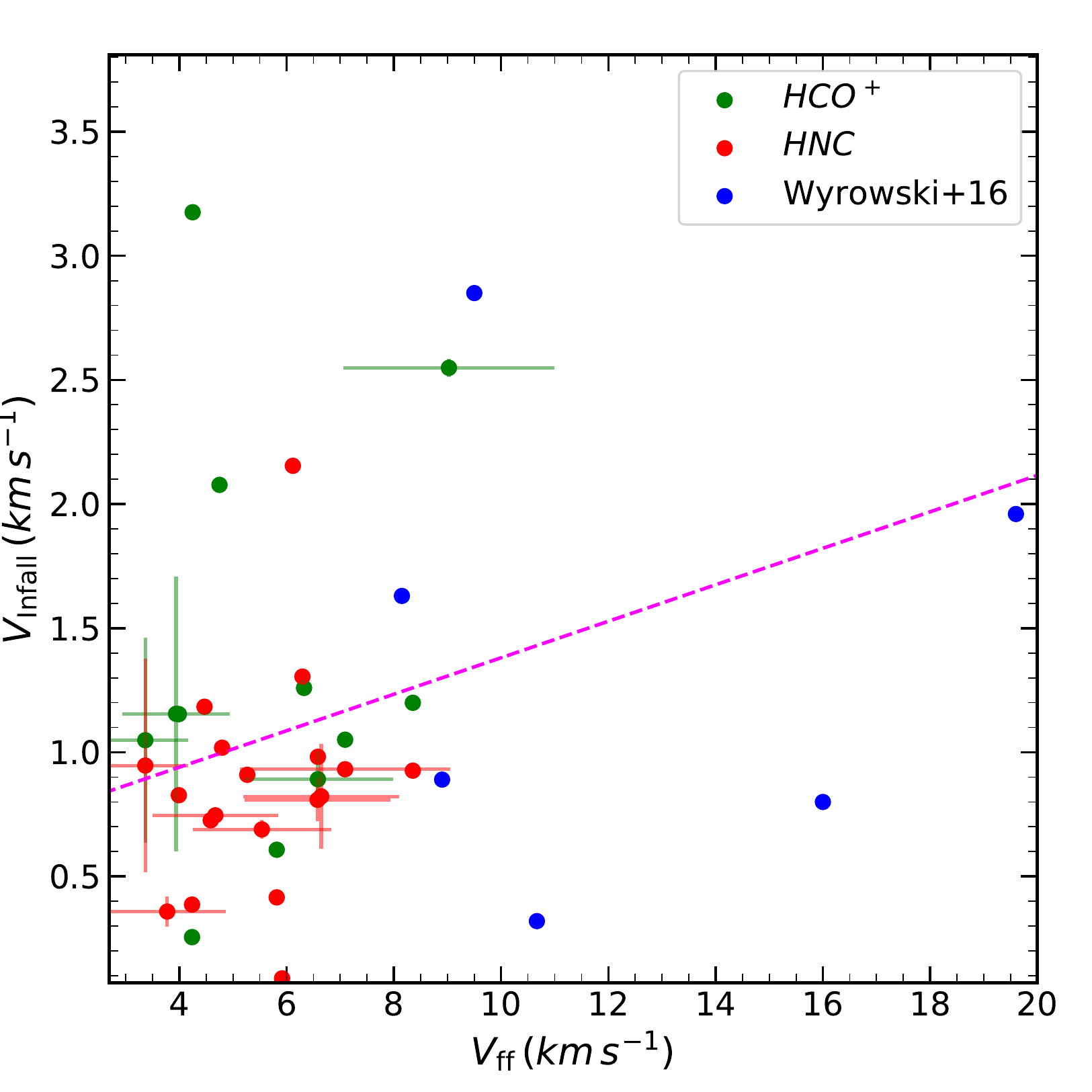} 
\caption{Comparison of the infall speeds as measured from the two optically thick lines with the free-fall velocity. The dashed line shows the results of a linear least squares fit to all of the point (slope = $0.07\pm0.08$, intercept = $0.65\pm 0.46$). The uncertainties are shown for every third source.}   \label{fig:infall_free_velocity}
\end{figure}

The free fall velocity can be calculated using:

\[
V_{\rm ff} = - \sqrt{\frac{2GM_{\rm enc}}{R}}
\]

\noindent where $M_{\rm enc}$ is the mass enclosed within a radius $R$. Using the FWHM clump mass and FWHM radius from \citet{urquhart2022} we have calculated this parameter for 75 clumps where these two parameters are available. In Figure\,\ref{fig:infall_free_velocity} we compare these free-fall speeds to the infall velocities determined by the model. There is no correlation ($r_s = 0.18$ with $p$-value = 0.32) between these two velocities and the slope returned by a linear least squares fit to the data is consistent with zero. With the exception of four sources all of the infall velocities are between 0.5 and 1.5\,\kms\ indicating that in many cases the infall speed is independent of the gas density. This would suggest that all clumps collapse at a similar rate perhaps limited by magnetic support and turbulence. It is also likely that the two approaches do not probe identical volumes of the same clump. While \citet{urquhart2022} provide the properties of the entire clump, our data provides us a velocity from a clump radius where the \HCOP\ optical depth is unity.  Differences in the radial structure of clumps may therefore dilute any correlation of these two different radii. Exploring either of these scenarios is beyond the scope of the current study but would be worth investigating further with a larger statistical sample. Note that have also included free-fall calculations and THz NH$_3$ absorption line measurements for high-mass star forming targets where the presence of inward motions has been established more directly \citep{wyrowski2016}. Those reference data exhibit the same trend seen in our observations, i.e., that inward motions are slower than free-fall collapse and would favor a scenario where infall may be regulated by magnetic field support and turbulence.

We use the infall-velocities determined from the HNC and \HCOP\ transitions to estimate the mass accretion rate for these clump using the following equation assuming that the clumps are spherical:

\[
\dot M = 4\pi R^2 \rho V_{\rm in},
\]

\begin{table*}

    \centering
     \caption{Infall velocities and mass infall rates for all sources where it has been possible to fit the blue asymmetric line profiles.}
    \begin{tabular}{ll.....}
\hline
ATLASGAL  & Classification & \multicolumn{1}{c}{$L_{\rm bol}/M_{\rm fwhm}$} & \multicolumn{1}{c}{HCO$^{+}\,V_{\rm in}$} & \multicolumn{1}{c}{$\dot M_{\rm HCO^{+}}$} & \multicolumn{1}{c}{HNC\,$V_{\rm in}$} & \multicolumn{1}{c}{$\dot M_{\rm HNC}$} \\

 Name  &  & \multicolumn{1}{c}{(L$_\odot$/M$_\odot$)} & \multicolumn{1}{c}{(\kms)} & \multicolumn{1}{c}{($10^{-3}$\,M$_\odot$\,yr$^{-1}$)}& \multicolumn{1}{c}{(\kms)} & \multicolumn{1}{c}{($10^{-3}$\,M$_\odot$\,yr$^{-1}$)} \\
\hline\hline
AGAL008.206+00.191	&	Protostellar	&	0.93	&	1.05	&	2.1	&	0.95	&	1.9	\\
AGAL008.706$-$00.414	&	Protostellar	&	0.58	&	1.20	&	14.7	&	0.93	&	11.4	\\
AGAL009.284$-$00.147	&	Protostellar	&	0.37	&		&		&	0.91	&	4.4	\\
AGAL009.879$-$00.111	&	Quiescent	&		&	0.85	&		&	0.68	&		\\
AGAL010.686$-$00.126	&	Protostellar	&	0.04	&		&		&	0.82	&	6.4	\\
AGAL011.381+00.811	&	Quiescent	&	0.21	&		&		&	1.02	&	4.1	\\
AGAL013.248+00.044	&	Quiescent	&	0.09	&	2.08	&	8.2	&		&		\\
AGAL013.906$-$00.292	&	Quiescent	&	0.32	&	1.16	&	3.2	&		&		\\
AGAL014.644$-$00.117	&	Quiescent	&	0.82	&	0.26	&	0.8	&	0.39	&	1.2	\\
AGAL014.726$-$00.202	&	Quiescent	&	0.28	&		&		&	1.43	&		\\
AGAL016.343+00.922	&	Protostellar	&	0.17	&	0.60	&		&	0.61	&		\\
AGAL016.418$-$00.634	&	Quiescent	&	0.22	&		&		&	0.81	&	6.2	\\
AGAL016.681$-$00.069	&	Protostellar	&	0.29	&		&		&	0.64	&		\\
AGAL019.394$-$00.006	&	Quiescent	&	0.3	&		&		&	1.18	&	4.2	\\
AGAL019.902$-$00.582	&	Quiescent	&	0.12	&	1.01	&		&		&		\\
AGAL022.056+00.191	&	Quiescent	&	0.54	&	1.15	&	3.2	&	0.83	&	2.3	\\
AGAL022.304$-$00.629	&	Quiescent	&	0.05	&		&		&	0.69	&	3.7	\\
AGAL022.376+00.447	&	Protostellar	&	0.59	&	0.89	&	6.8	&	0.98	&	7.5	\\
AGAL022.726+00.127	&	Quiescent	&	0.22	&	1.23	&		&	1.20	&		\\
AGAL023.477+00.114	&	Quiescent	&	0.38	&		&		&	1.30	&	9.1	\\
AGAL024.049$-$00.214	&	Quiescent	&	0.13	&	1.26	&	8.9	&		&		\\
AGAL024.089$-$00.104	&	Quiescent	&	0.61	&		&		&	0.36	&	0.9	\\
AGAL024.373$-$00.159	&	Protostellar	&	0.38	&	3.18	&	10.1	&		&		\\
AGAL024.378$-$00.209	&	Quiescent	&	0.34	&	1.67	&		&	1.35	&		\\
AGAL024.574$-$00.074	&	Protostellar	&	0.53	&		&		&	0.93	&	2.0	\\
AGAL028.374+00.054	&	Quiescent	&	0.16	&		&		&	2.15	&	14.2	\\
AGAL030.718+00.191	&	Quiescent	&	0.04	&	0.86	&		&		&		\\
AGAL030.844+00.177	&	Quiescent	&	1.15	&		&		&	0.73	&	2.7	\\
AGAL030.913+00.719	&	Protostellar	&	0.27	&		&		&	0.75	&	2.9	\\
AGAL031.699$-$00.494	&	Quiescent	&	0.08	&		&		&	0.09	&	0.6	\\
AGAL031.946+00.076	&	Protostellar	&	0.06	&	2.55	&	36.5	&		&		\\
AGAL035.431+00.137	&	Quiescent	&	0.64	&	0.61	&	3.6	&	0.42	&	2.5	\\
AGAL035.479$-$00.309	&	Quiescent	&	0.05	&	1.05	&	9.3	&	0.93	&	8.2	\\
\hline

    \end{tabular}
   
    \label{tab:infall_rate}
\end{table*}

\noindent where $V_{\rm in}$ is the infall velocity determined by the model, $R$ is the radius and  $\rho$ is the
clump density, which can also be written as 

\[
\rho = \frac{3M}{4\pi R^3},
\]

\noindent where $M$ is the clump mass. Converting the mass to solar masses, radius to parsecs and the infall velocity to \kms\ the infall mass becomes:

\[
\dot M = 3.066\times 10^{-6} \times \left(\frac{M}{\rm M_\odot}\right)\, \left(\frac{R}{\rm pc}\right)^{-1}\, \left(\frac{V_{\rm in}}{\rm km\,s^{-1}}\right)\, \left[\rm M_\odot\,yr^{-1} \right].
\]

We have used the FWHM clump mass and radius from \citet{urquhart2022} and the infall velocities determined above to calculate the mass infall rates. {These are given in Table\,\ref{tab:infall_rate} and are between 0.6 and $36\times 10^{-3}$\,M$_\odot$\,yr$^{-1}$. If instead, the \HCOP\ probed gas volume is different from that probed by ATLASGAL based dust emission, the mass infall rates have to be revised accordingly. We assume a characteristic volume density of $\sim 1000$\,\percc\ for the \HCOP\ 1-0 line \citep{kauffmann2017}.  We also need to extract a radius for the typical \HCOP\ emission. For a 10-20\,K dust clump, all of the ATLASGAL based dust emission will be above a H$_2$ column density of $2 \times 10^{22}$ \cmsq{}, significantly above the threshold of dense gas mass residing at $A_V>7$\,mag \citep{lada2010}.  Therefore, \HCOP\ emitting region should extend beyond the typical 30-60\,\arcsec\ radius from dust measurements (e.g. \citealt{hoq2013, miettinen2014}). Assuming a 1\,parsec radius ($\sim 70$\,\arcsec\ at 3\,kpc), we then derive mass infall rates that are between 0.4 and $3\times 10^{-3}$\,M$_\odot$\,yr$^{-1}$. Though the latter approach may be more imprecise, both estimates are consistent with the ranges reported by previous studies of high-mass star forming clumps via absorption spectroscopy (e.g. $0.3-16 \times 10^{-3}$\,M$_\odot$\,yr$^{-1}$; \citealt{wyrowski2016}) or similar approaches as used here (e.g. $0.7-71 \times 10^{-3}$\,M$_\odot$\,yr$^{-1}$; \citealt{he2015}, $0.5-45 \times 10^{-3}$\,M$_\odot$\,yr$^{-1}$; \citealt{traficante2018b}).

\subsection{SiO Outflows}

Another spectroscopic approach to distinguishing protostellar clumps from starless or pre-stellar clumps is via searches for molecular outflows (e.g. \citealt{maud2015,de_villiers2014,csengeri2016_sio, yang2018}). While CO is the main outflow tracer, low resolution observations in the inner galactic plane particularly of the low excitation (J=1-0, 2-1) transitions are often contaminated by ambient cloud emission as well as by unrelated diffuse and dense emission along the line of sight. SiO is believed to form through sputtering and grain-grain collisions of dust grains (\citealt{schilke1997}). Unlike CO, SiO emission does not suffer from contamination from easily excited ambient gas or unrelated line of sight components. While SiO emission with a narrow velocity range may originate from large-scale colliding gas flows (e.g., \citealt{jimenez-Serra2010}, \citealt{cosentino2020}), SiO emission with a broad velocity range is considered to be an effective  tracer of fast shocks from protostellar outflows. 

In this work, we also search for broad SiO emission as an indirect tracer of outflows from deeply embedded protostars. As described above, the related observations were obtained simultaneously with those of the infall tracers. We do the outflow search in the following manner. We inspected the spectrum of each source close to the rest frequency of the SiO (2-1) transition, and consider the line to be detected if the peak intensity in the spectrum is larger than 5$\sigma$ (where $\sigma$ corresponds to the rms noise per velocity channel). We then fit the hyperfine structure of the 1--0 transition of \NTH\ to determine the characteristic dense gas line widths. Line widths for this sub-sample are in the range 1.2--3.2\,\kms\ with a mean value of 2.2$\pm0.5$\,\kms. The maximum line width (3.2\,\kms) is used as a threshold to separate broad and narrow SiO components. All sources have line widths that exceed that of \NTH\ and 90\,per\,cent have line widths that are larger by at least a factor two. Therefore we consider these to be bonafide outflow candidates.
The SiO emission towards thirty-one clumps satisfy this criterion, corresponding to a low detection rate of 28\,per\,cent. This is consistent with the general trend gleaned from previous outflow searches towards ATLASGAL sources that at least on the clump scales,  the outflow signatures in the earlier stages are weaker \citep{yang2022, csengeri2016_sio}.

\section{Discussion}
\label{sect:discussion}

\subsection{Mass infall rate and protostellar evolution}
\label{sect:evolution_of_mass_infall_rate}

As mentioned previously, there is some evidence that the infall motion is lower for more evolved protostellar stages (\citealt{jackson2019}) suggesting that the mass infall rate decreases as the embedded pro-cluster evolves.  Our sample contains few clumps in the later evolutionary stages, so we are unable to conduct a similar analysis. However, we can look for changes in the mass infall rate as a function of the luminosity-to-mass ratio ($L/M$), which is considered to be a good diagnostic for the state of star formation within a clump with larger values indicating more advanced evolution (e.g. \citealt{molinari2008,urquhart2014_atlas}). 

In Figure\,\ref{fig:infall_vs_LM_ratio} we show the distribution of the mass infall rate as a function of the bolometric luminosity and FWHM clump mass ($L_{\rm bol}/M_{\rm fwhm}$; \citealt{urquhart2022}). This plot reveals a negative correlation between these two parameters with the mass infall rate decreasing with increasing evolution, which is consistent with the trend reported by \citet{jackson2019}. However,  the correlation coefficient is $r_s = -0.36$ with a $p$-value = 0.045, indicating a weak correlation with the significance just above 2$\sigma$ and so more data is needed to robustly confirm this trend.

\begin{figure}
  \centering    
  
  \includegraphics[width=.48\textwidth]{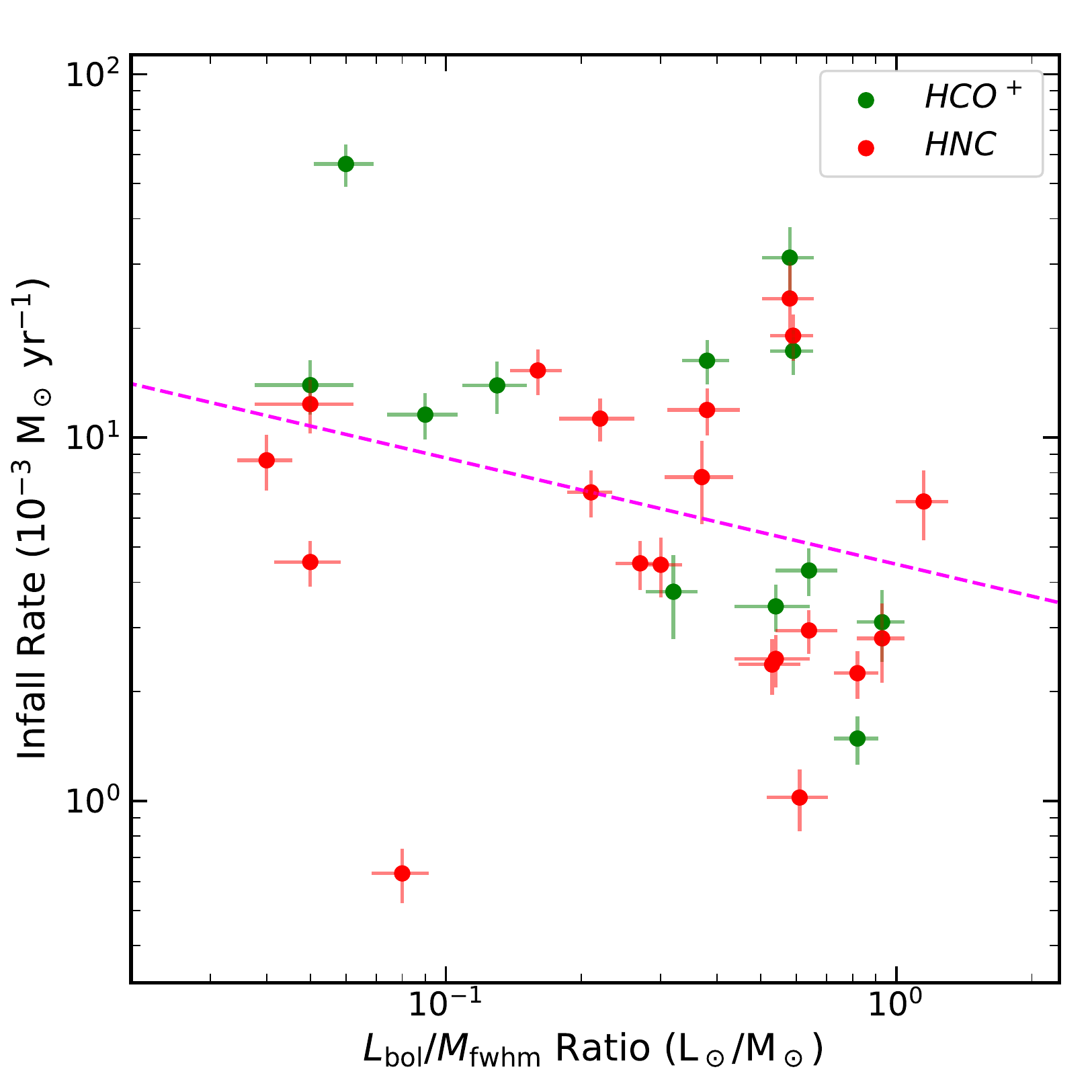} 
\caption{Mass infall rates as a function of the $L/M$ ratio. The values determined from the two transition are shown in different colours and the dashed line shows the results of a linear least squares fit to the log values of both parameters (slope = $-0.29\pm0.18$, intercept = $0.65\pm 0.12$). }   \label{fig:infall_vs_LM_ratio}
\end{figure}

\subsection{Star formation in quiescent clumps}

\begin{table}

    \centering
     \caption{Infall and outflow statistic for the quiescent and protostellar clumps. \label{tbl:infall_outflow_stats}}
    \begin{tabular}{lcccc}
\hline
Type  & \# of clumps & Infall & Ouflow & Total \\
\hline\hline
Quiescent & 70 & 42 (60\%) & 14 (20\%) & 47 (67\%)\\
Protostellar & 23 & 17 (74\%) & 13 (57\%) & 20 (87\%) \\
\hline
\end{tabular}
\end{table}

One of the main aims of this work was to investigate whether there is evidence of gravitational collapse or star formation activity towards a sample of clumps in the earliest stages in their evolution. In Table\,\ref{tbl:infall_outflow_stats} we present a summary of the infall and outflow associations for the quiescent and protostellar clumps. 

Almost two-third of the sample has been classified as quiescent and a further 20\,per\,cent are classified as being protostellar. Of the 70 quiescent sources in the sample, 42 are associated with infall motion (corresponding to 60\,per\,cent of the sample), 22 of which we have been able to measure the infall velocity for. In addition to the infall motions detected, we have detected SiO emission towards 31 clumps including 14 of the quiescent clumps. In total, 47 of the clumps classified as being quiescent show evidence of either infall or have been identified as outflow candidates, or both, which corresponds to $\sim$67\,per\,cent of the quiescent sample (i.e. two-thirds).

The fraction of quiescent sources associated with SiO emission is 20\,per\,cent, which is less than half of the association rate found for the protostellar sample (57\,per\,cent), suggesting that the incident of outflows increases with evolution (e.g. \citealt{urquhart2022}). The number of protostellar sources associated with infall motion is 74\,per\,cent. This is again significantly higher than found for quiescent clumps where only 57\,per\,cent are associated with infall motions. Approximately 87\,per\,cent of protostellar clumps are associated with either infall or outflow motions, which is unsurprising given their nature. 

\citet{yang2022} recently conducted an outflow survey towards $\sim$2000 ATLASGAL clumps using the SEDIGISM survey (\citealt{schuller2021}) and report approximately 50\,per\,cent of the quiescent and protostellar clumps are associated with outflows. Their detection rate towards protostellar clumps is comparable with ours but significantly higher than we have found for the quiescent clumps. However, Yang et al. have used the 2-1 transition of $^{13}$CO, a molecule whose abundance even in typical molecular cloud conditions is much higher than that of SiO \citep{cosentino2020}. Only 13 targets from our sample overlap with the Yang et al. sample, out of which 6 have $^{13}$CO associations and 4 of these also have have been detected in our SiO observations. Two sources identified as SiO outflow candidates have no counterpart in $^{13}$CO.
This suggests that more of the quiescent sources may be associated with outflows and our detection statistic is likely to be a lower limit.

Although the sample examined here is relatively modest in size (110) it does reveal a couple of interesting trends. First, that a significant number of the clumps classified as be being quiescent  appear to be undergoing gravitational collapse with a smaller fraction being associated with SiO emission, which suggests that many of these seemingly starless cores are pre-stellar in nature and at least in some of them protostars have already started to form despite the lack of a corresponding embedded infrared detection. Second, that the incidence rates of infall and outflow motions increase in the protostellar stage, consistent with the hypothesis that star formation in clumps classified as protostellar is more evolved.

\subsection{Towards an ALMA sample}

Although  a significant number of the quiescent clumps identified by ATLASGAL appears to be undergoing gravitational collapse, it is likely to be in a very early stage where the initial conditions unlikely to have been significantly affected by feedback. Furthermore, a significant number still appear to be genuinely quiescent  ($\sim$32\,per\,cent of the sample). This sample of quiescent clumps is, therefore, likely to include examples of all of the very earliest stages in a clumps evolution, from the contraction of the clumps themselves, fragmentation and formation of starless cores, and the formation of protostellar objects. 

Understanding the properties of these clumps and thus, the initial conditions in the earliest phase, is one of the most important steps that will lead us to a better understanding of the process of formation of a high-mass protostar and its subsequent evolution. However, while the single dish data discussed here provides a physical characterization of the high-mass clumps, only high angular resolution continuum and line data can characterize these candidates as true pre-stellar massive clumps. This has motivated us to identify a large and statistically representative sample of quiescence clumps to observe with ALMA.

The ATLASGAL sample is one of the most well characterised samples of high-mass star forming clumps available. It has identified $\sim$1000 quiescent and a similar number of protostellar clumps it is therefore an ideal starting point for selecting a large and representative sample of quiescent clumps. We have applied our selection criteria (described in Sect.\,\ref{sect:source_selection}) to the whole ATLASGAL catalogue (\citealt{contreras2013,urquhart2014_csc}) and have identified a sample of 238 quiescent clumps within 5\,kpc that satisfy the empirical mass-size threshold for HMSF \citep{kauffmann2010c, urquhart2014_atlas}; we refer to this sample as the \underline{Co}ld \underline{Co}res with \underline{A}LMA (CoCoA) sample. 

Given the similarities in selection criteria we expect the distribution of evolutionary stages to also be similar to that of the sample presented here (see Fig.\,\ref{fig:pie_chart} for details). The inclusion of some later stages is useful for comparison of the statistical properties of the quiescent clumps in an evolutionary framework in an identical observational setup.   This sample has subsequently been followed-up with ALMA and the first results of CoCoA survey will be presented in a subsequent paper (Pillai et al. 2023, in prep.).

\section{Conclusions}
\label{sect:conclusions}

We have observed a sample of 110 high-mass clumps at frequencies between 86 and 93\,GHz using the IRAM 30-m telescope. This sample consists primarily of clumps that have been classified as quiescent or in the protostellar stage (\citealt{urquhart2022}). This work uses the \HCOP\ and HNC (1-0) transitions to investigate infall within the clumps, and SiO (2-1) to identify potential outflow candidates. These are our main findings:

\begin{itemize}
    \item  We have conducted a statistical analysis to assess the presence of contraction and expansion motions in our data. This analysis evaluates, for every individual target, the uncertainty of the asymmetry measurement using two different methods, and then compares the observed asymmetry to its uncertainty. We find that the number of sources with line asymmetries exceeds the number expected from the impact of noise. Line asymmetries are thus substantial in our sample, indicating the presence of significant relative gas motions in our targets.\\

    \item We have detected SiO emission towards 31 clumps corresponding to $\sim28$\,per\,cent of the sample. These are considered to be good outflow candidates. We find that the detection rate is a factor of two higher for protostellar clumps than quiescent clumps, which is consistent with the star formation in the protostellar clumps being more evolved.\\
    
    \item Combining the infall and outflow tracers we find that 67\,per\,cent of the clumps classified as being quiescent are associated with the early signposts of gravitational collapse or star formation. Studying the star formation taking place in these clumps is likely to provide us with our best opportunity to determine the initial conditions required for the onset of star formation and determine how fragmentation proceeds and when the first protostars start to appear.\\
    
    \item We provide an overview of a systematic high-resolution ALMA study that has been designed to investigate the star formation in a large sample of quiescent ATLASGAL clumps. These observations will allow us to resolve these clumps on core scales and investigate the fragmentation statistics and derive the physical and star forming properties of the cores identified. This will reveal the initial conditions for high-mass star formation and allow us to develop a detailed understanding of earliest stages and subsequent evolution. \\

\end{itemize}

\section*{Acknowledgements}

TGSP gratefully acknowledges support by the National Science Foundation under grant No. AST-2009842 and AST-2108989. SL acknowledges support by the INAF PRIN 2019 grant ONSET. KW acknowledges science research grants from the China Manned Space Project (CMS-CSST-2021-A09, CMS-CSST-2021-B06), the National Key Research and Development Program of China (2017YFA0402702), and the National Science Foundation of China (12041305, 11973013). The authors would like to thank the reviewer for their constructive and insightful comments. The authors would like to thank Jim Jackson and Scott Whitaker for making their published data on asymmetry parameters of the MALT90 sample available upon request. 

\section*{Data Availability}
The data underlying this article will be shared on reasonable request to the corresponding author.


\bibliographystyle{mnras}
\bibliography{urquhart_2021}

\end{document}